\def \pbpb  {Pb+Pb}
\def \pp    {$pp$ }
\def\corr{\langle \Delta p_{t,1}, \Delta p_{t,2}\rangle}

\RequirePackage{lineno}
\documentclass[amsmath,amssymb]{revtex4}

\usepackage{epsfig}
\usepackage{graphicx}
\usepackage{dcolumn}
\usepackage{bm}
\usepackage{setspace}

\begin{document}

\title{Heavy-quark azimuthal momentum correlations as  a sensitive probe of
thermalization }


\author{G.~Tsiledakis$^1$, H. Appelsh\"auser$^2$, K.~Schweda$^1$, J.~Stachel$^1$
}
\affiliation{$^1$Physikalisches Institut der Universit\"at Heidelberg, Philosophenweg 12,\\ D-69120 Heidelberg, Germany\\
$^2$Institut f\"ur Kernphysik der Universit\"at Frankfurt,\\ D-60438 Frankfurt, Germany
}%


\date{\today}

\begin{abstract}
In high-energy nuclear collisions 
the degree of thermalization at the partonic level is a key issue. 
Due to their large mass,
heavy quarks and their possible participation in the collective flow of the QCD-medium
constitute a powerful probe for thermalization.
We present studies with PYTHIA for
$pp$ collisions at the top LHC energy of $\sqrt{s}$ = 14 TeV applying the two-particle transverse momentum correlator 
$\corr$ to pairs of  heavy-quark hadrons
and their semi-leptonic decay products as a function of their relative azimuth.
Modifications or even the complete absence of initially existing correlations in \pbpb\ 
collisions might indicate thermalization at the partonic level. 
\end{abstract}

\pacs{Valid PACS appear here}
\maketitle

 \section{Introduction}
High-energy nuclear collisions offer the unique opportunity to probe highly
excited nuclear matter in the laboratory. At sufficiently high
temperature and/or energy density hadrons dissolve 
and quarks and gluons carrying color charge
become the relevant degrees of freedom. This state of matter is commonly called
a Quark Gluon Plasma (QGP).
An essential difference between collisions of leptons or hadrons  on the one hand
and heavy nuclei on the other hand is the development of collectivity in the
latter. Collective flow of hadrons, especially the strange hadrons $\phi$
and $\Omega$, has been experimentally observed at the Relativistic Heavy-Ion Collider (RHIC)~\cite{strange} suggesting that collectivity
significantly develops in the early partonic stage, i.e. among quarks and gluons. Presently, the degree of
thermalization among partons remains a key issue.

Heavy-quark hadrons and their observables are of particular interest when addressing
thermalization~\cite{thermalization}. In a QGP, chiral symmetry might be partially restored 
and in that case quark masses should approach the current quark masses. 
While these are small compared to the temperature $T$ for up and down quark and comparable
to $T$ for the strange quark, they are large compared to $T$ for the heavy quarks (charm and beauty)~\cite{quark-masses}.
Therefore heavy quarks are mostly created in the early stage of the collision.
Further, annihilation of heavy quarks in the QCD-medium is negligible~\cite{pbm_2007}. 
Thus, heavy quarks probe the entire history of a high-energy nuclear collision.
Heavy quarks participate in collective motion provided
interactions at the partonic level occur frequently. In general, frequent interactions 
drive a system towards local thermal equilibrium.
Thus, collective motion
of heavy-quark hadrons is a powerful tool when addressing early thermalization of
quarks and gluons in high--energy nuclear collisions.

In strong interactions
heavy quarks are always created together with their anti--quark and are thus correlated.
In collisions of leptons it has been shown that a hadron containing a charm or beauty quark carries a 
significant fraction of the initial quark momentum~\cite{aleph_95,delphi_95,opal_95}.
Hence, the initial heavy-quark correlations survive the fragmentation
process into hadrons to a large extent and are observable e.g. in the angular distributions 
of pairs of $D$-- and $\overline{D}$--mesons ~\cite{review,e791}. 

In high--energy collisions of heavy nuclei, frequent interactions among partons (quarks and gluons) of the 
medium and heavy quarks
may lead to a significant modification of these initially existing correlations.
On the other hand, hadronic interactions at the late stage are insufficient
to alter azimuthal correlations of 
$D\overline{D}$--pairs~\cite{Zhu}. 
Frequent interactions
distribute and randomize the available (kinetic) energy and finally drive the system, 
i.e. light quarks and gluons,
to local thermal equilibrium.
To what extent this also happens for heavy quarks is currently a subject of discussion~\cite{hq-thermalization}.
An experimental tool to address this question is studied in the current publication.   
A decrease in the strength of heavy quark correlations in high--energy collisions of heavy nuclei 
as compared to \pp collisions
would indicate early thermalization also of heavy quarks.

We have employed the Monte Carlo event generator PYTHIA \cite{pythia} which
implements a fragmentation scheme model that 
reproduces 
experimentally observed correlations of $D$--mesons at fixed target energies~\cite{review}.
To extend calculations in PYTHIA beyond leading order, processes contributing at higher orders
were calculated using a massless matrix element~\cite{ALICE_PPRII} applying 
a lower cut-off in the transverse momentum-transfer scale of the underlying hard scattering to avoid divergences in
the calculated cross section~\cite{ALICE_PPRII}.   
The PYTHIA parameters were subsequently tuned to reproduce these next-to-leading order predictions~\cite{ALICE_PPRII}.
For the hadronization of charm and bottom quarks 
the Lund fragmentation scheme was used~\cite{pythia2}.  

Our calculations at leading order (LO)  contain flavor
creation processes ($q\overline{q}\rightarrow Q\overline{Q}$, $gg\rightarrow Q\overline{Q}$)
and lead to an enhancement at relative azimuth around $\Delta\phi \approx 180^{\rm o}$ , i.e. 
the $D$--meson pair is preferentially emitted back-to-back. 
Here, lower-case letters denote light quarks ($q$) and gluons ($g$) while upper-case letters denote
heavy quarks and anti-quarks ($Q,\overline{Q}$).
Next-to-leading order (NLO) contributions such as flavor excitation
($qQ\rightarrow qQ, gQ\rightarrow gQ$) and gluon splitting ($g\rightarrow Q\overline{Q}$) 
have a strong dependence on the center-of-mass energy and become dominant at LHC. 
It was noted in~\cite{Zhu2}  that
these higher-order processes do not show pronounced 
features in azimuth and weaken or destroy the azimuthal correlation between charmed hadrons.
By choosing momentum cuts, a somewhat enhanced correlation was extracted~\cite{Zhu2}. 

With these processes expected to dominate at LHC energies, sensitivity to heavy quark thermalization
might be lost. We investigate whether the  $p_{t}$ correlator as a more sensitive measure 
of correlations would again provide sensitivity to thermalization.
We introduce
the two-particle transverse momentum correlator as a sensitive measure of heavy-quark correlations. 
This method has the following advantages: \newline
(i) The correlator is sensitive to non-statistical fluctuations, thus carving out any physical correlation.\newline
(ii) In case of physically uncorrelated candidate--pairs (e.g. background), the extracted value for the correlator vanishes,
thus providing a reliable baseline.\newline
(iii) A localization of the observed correlations in transverse momentum space may help to obtain further insight into
the origin of the observed correlations in relative azimuth.

\section{Employing the two-particle transverse momentum correlator}
\begin{figure}[h]
\centering
\includegraphics[width=0.49\textwidth]{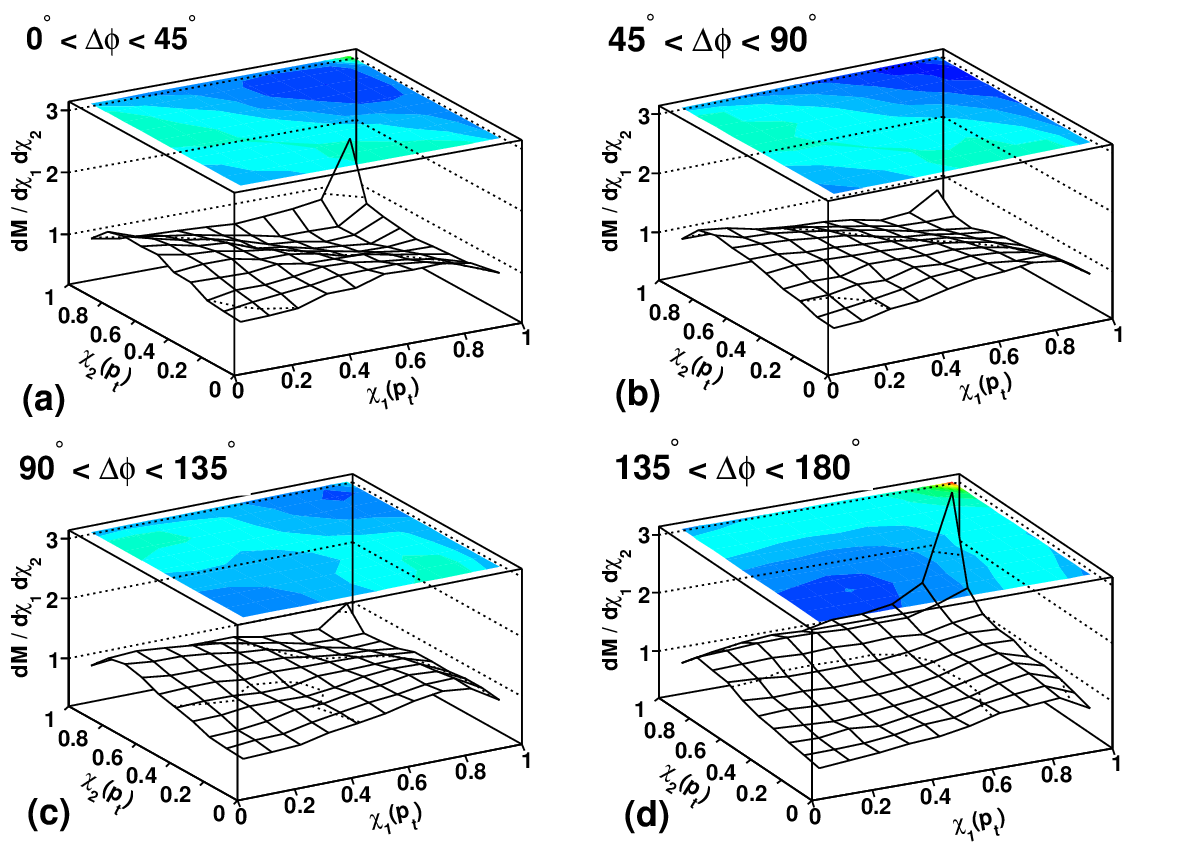}
\caption{(Color online) Two-particle correlations as function of $(\chi(p_t)_1, \chi(p_t)_2)$ of 
500k $D\overline{D}$ pairs in different regions of $\Delta\phi$ 
integrated over full rapidity for $pp$ collisions at $\sqrt{s}$ = 14 TeV as calculated using PYTHIA (v. 6.406). }
\label{cumul}
\end{figure}

We studied two-particle correlations of 
$D\overline{D}$--pairs in the 
$(\chi(p_t)_1, \chi(p_t)_2)$-plane, with the cumulative variable $\chi(p_t)$ defined as 
 \cite{ptcumul,na49PRL}:
  \begin{equation}
 \chi(p_{t})=\int\limits_{0}^{p_{t}}\rho(p_{t}^{'}) dp_{t}^{'}.
 \end{equation}
 Here, $\rho(p_{t}^{'})$ is the inclusive $p_{t}$ distribution, normalized to unity, which is 
 obtained from all $D\overline{D}$--pairs used in the analysis. 
 For the study of two-particle $p_{t}$ correlations, the   
$\chi(p_t)$-values of $D\overline{D}$--pairs $(\chi(p_t)_1, \chi(p_t)_2)$ are filled into two-dimensional arrays.
In Fig.~\ref{cumul}, the two-particle correlation function $\frac{dM}{d\chi_{1} d\chi_{2}}$ is shown for different values of the 
azimuthal separation $\Delta\phi = \phi_{D} - \phi_{\overline{D}}$ of the $D\overline{D}$--pair.
For small values in $\Delta\phi$ we observe a strong positive correlation 
at $\chi_1 \approx \chi_2 \approx 1$. This correlation is most
pronounced in the high $p_{t}$-region related to gluon splitting processes.
 At large values of $\Delta\phi
\approx 180^{\rm o}$, a
substantial positive correlation in the high $p_{t}$-region comes from flavor creation
processes. 

The occurrence of non-statistical fluctuations of the event-by-event mean 
transverse momentum $M_{pt}$ goes along 
with correlations among the transverse momenta of particle pairs~\cite{koch99}. 
Such correlations were successfully extracted from experimental data
employing the two-particle transverse momentum correlator~\cite{ceres-pt,spt,spt1,spt2,spt3}.
For the study of $p_t$ correlations between particles of different
charge sign like $D$ and $\overline{D}$ mesons, the
correlator is calculated in the following way:
\begin{equation}
\corr^{(D\overline{D})} =
\frac{1}
{\sum_{k=1}^{n_{\rm ev}}N_k^{D}N_k^{\overline{D}}}.
\sum_{k=1}^{n_{\rm ev}}
\sum_{i=1}^{N_k^{D}}\sum_{j=1}^{N_k^{\overline{D}}}(p_{ti}-\overline{p_t}^{(D)})
(p_{tj}-\overline{p_t}^{(\overline{D})}) 
\end{equation}
where $p_{ti}$ and $p_{tj}$ are the transverse momentum of the $i^{th}$ and $j^{th}$ 
$D$-- and $\overline{D}$--meson.
Here, the index $i$ runs over all particles, while the index $j$ runs over all anti--particles created in a single \pp collision.
The inclusive mean transverse momentum is 
averaged over all  $D$ and $\overline{D}$--mesons, respectively, and
is denoted by $\overline{p_t}$.  
The total number of 
$D\overline{D}$ pairs summed over the number of \pp collisions $n_{ev}$ is given by
 $\sum_{k=1}^{n_{\rm ev}}N_k^{D}N_k^{\overline{D}}$, with $N_k^{D}$ and $N_k^{\overline{D}}$ the number of $D$ and $\overline{D}$ mesons created in a single  \pp collision. Note that in
the present studies we generated 
one $D\overline{D}$ pair per \pp collision,  $N_k^{D} = N_k^{\overline{D}} = 1$.    
In total, we generated 2M (500k) \pp collisions with a $D\overline{D}$ ($B\overline{B}$) pair.

We studied $p_{t}$ correlations 
and their dependence on azimuthal separation by calculating the correlator in bins of 
$\Delta\phi$.
In case of independent particle emission, the correlator $\corr$
vanishes. 
\begin{figure}[bp]
\centering
\includegraphics[width=0.49\textwidth]{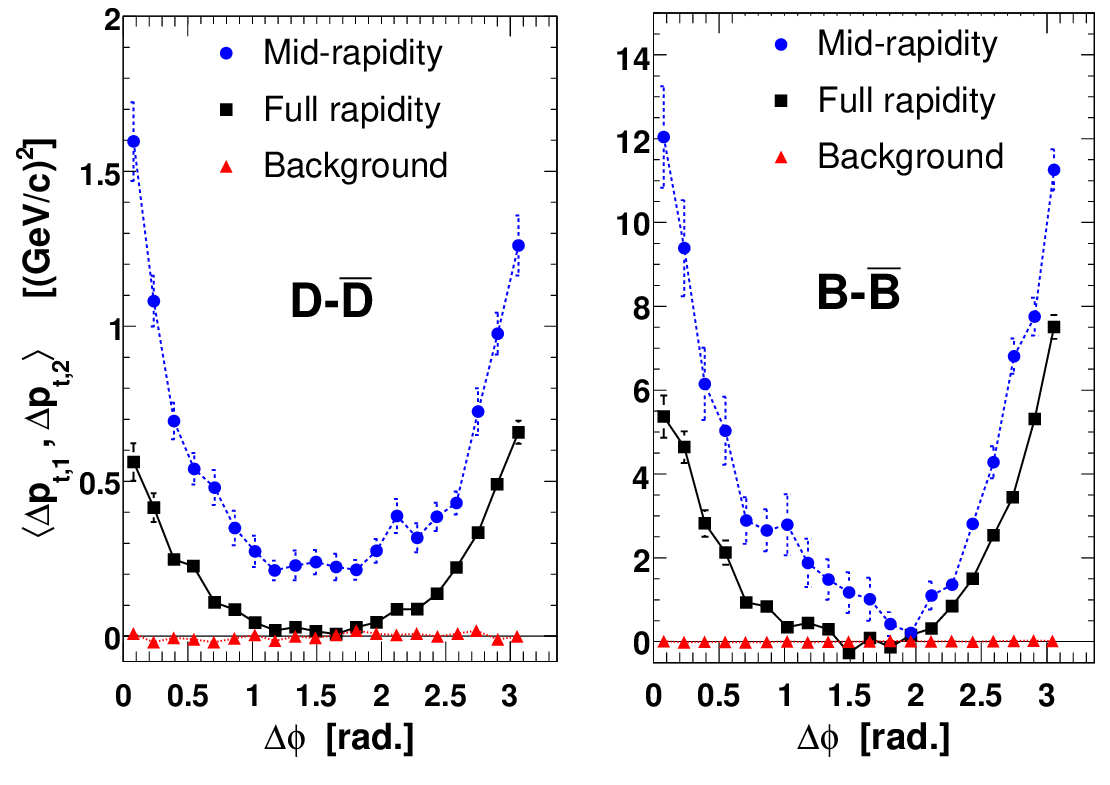}
\caption{(Color online) Distribution of  the momentum correlator $\corr$ of 500k $D\overline{D}$ pairs (left panel) and 
500k $B\overline{B}$ pairs (right panel) 
as a function of relative azimuth $\Delta\phi$ 
at mid-rapidity~(circles), for full rapidity~(squares) and for background using the mixed event method (triangles) 
for $pp$ collisions at $\sqrt{s}$ = 14 TeV as calculated using PYTHIA (v. 6.406). The lines connect the points. }
\label{DDbarBBbar}
\end{figure}
\begin{figure}[h]
\centering
\includegraphics[width=0.49\textwidth]{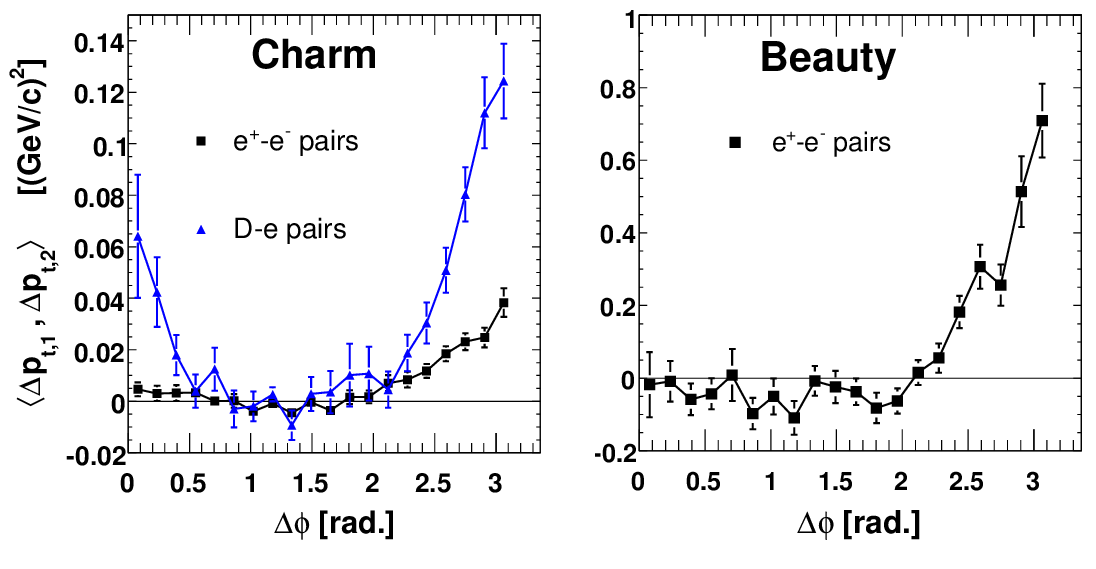}
\caption{(Color online) Distribution of  the momentum correlator $\corr$ of 200k $e^{+}e^{-}$-pairs from $D\overline{D}$ 
decays and $De$ correlations (left panel) and of 50k $e^{+}e^{-}$-pairs from $B\overline{B}$ decays (right panel) as a function of relative azimuth $\Delta\phi$ 
for full rapidity
for $pp$ collisions at $\sqrt{s}$ = 14 TeV as calculated using PYTHIA (v. 6.406). The lines connect the points. }
\vspace{-5mm}
\label{DeBe}
\end{figure}
The $D\overline{D}$ momentum correlator $\corr$ from
our simulations of \pp collisions at $\sqrt{s}$ =~14~TeV  
is shown in the left panel of Fig.~\ref{DDbarBBbar} 
as a function of $\Delta\phi$.  The error bars reflect the statistical uncertainties from our finite data sample.
The correlator has a pronounced forward-backward peaked structure.
We observe an enhancement at small azimuth from gluon
splitting processes, while flavor creation of $c\overline{c}$--quark pairs leads to an enhanced  correlation at backward angles.
We have checked that flavor excitation processes, involving a larger number of gluons, lead to a rather flat distribution. Also, our studies show that the correlations are even stronger
at mid-rapidity when compared to full rapidity which can be attributed to the harder particle spectrum at mid-rapidity.
Integrating the correlator over all azimuth and full rapidity, we get
$\corr = 0.199\pm0.006$~GeV$^2$/$c^2$.

In order to account for a possible change in the single particle spectrum when comparing different collision systems or energies, the normalized  dynamical 
fluctuation 
 \begin{equation}
\Sigma_{pt} =  {\rm sgn}(\corr) \frac{\sqrt{|\corr|}}{\overline{p_t}}
 \end{equation}
has been introduced as a dimensionless measure~\cite{spt}.

Our result for $D\overline{D}$ mesons as given above
amounts to $\Sigma_{pt} \approx 28\%$ 
with $\overline{p_t}^D$ = 1.58~GeV/$c$. 
This represents a large value implying a strong
correlation when compared to e.g. 
$\Sigma_{pt} \approx 1\%$ observed for unidentified charged particles in central collisions at SPS and RHIC~\cite{spt,spt1,spt2,spt3}. 
To mimic combinatorial background which is always present in the experiment, we applied the correlator 
to  $D$-- and $\overline{D}$-- mesons  from different \pp collisions, which are physically uncorrelated. 
This results in a value of $\corr$ consistent with zero (see Fig.~\ref{DDbarBBbar}). Therefore the correlator
allows for a clear distinction between the case were correlations are present (different from zero) or absent (equal to zero)
going beyond the method described in~\cite{Zhu, Zhu2}. 
%

The $B\overline{B}$ momentum correlator is shown in the right panel of
Fig.~\ref{DDbarBBbar} and has a structure
similar to the one for $D\overline{D}$--pairs.
Integrating the correlator over all azimuth and full rapidity, we get
$\corr = 2.73\pm0.05$~GeV$^2$/$c^2$  which corresponds to the normalized  
fluctuation $\Sigma_{pt} \approx 31\%$
with $\overline{p_t}^B$ = 5.27~GeV/$c$. 
This demonstrates that by applying the momentum correlator to  pairs of heavy quarks in 
\pp collisions at LHC energies, strong correlations are predicted which should be experimentally observable. When only considering 
$D\overline{D}$ production yields as a function of relative azimuth, a weaker dependency is predicted~\cite{Zhu2}.  

At LHC energies, the production of $D$--mesons is dominated by gluon splitting and flavor excitation processes while the contribution
from flavor creation is about 10\% at low momentum and increases up
to 20\% at larger momentum~\cite{sqm08}. 
On the other hand, the production of $B$--mesons is dominated by flavor creation and flavor excitation with a small contribution
 below 10\% from gluon splitting and an overall weak dependence on transverse momentum~\cite{sqm08}. 

As shown above, the initial correlations of $c\overline{c}$--quark pairs survive the fragmentation process into hadrons to a large extent. 
However experimentally, full kinematic reconstruction of $D$--mesons from topological decays suffer from small branching ratios and 
rather small reconstructing efficiencies resulting in low statistics,
especially when pairs of  $D$--mesons are considered where these factors enter quadratically.
In minimum bias \pp collisions at $\sqrt{s}$ = 14 TeV, roughly 28 (2) out of 1000 collisions create 
a charmed (beauty) meson such as $D^0, D^+$ or $ D_s^+$  ($B^0, B^+$ or $B_s^0)$ at mid-rapidity $|y| < 1$. 
The branching ratio in the golden channel $D^0 \rightarrow K^- + \pi^+$ amounts to $3.83 \%$ with additional penalty factors
due to the detector acceptance for the decay particles and topological reconstruction of the secondary decay vertex.
Overall we estimate the number of fully reconstructed $D^0-\overline{D^0}$ pairs in $10^9$ minimum bias \pp collisions, which is
equivalent to one year of ALICE data taking, to be in the order of 10. This is obviously too low to study $p_{t}$ correlations.

As an alternative, we considered electrons (positrons) from semi-leptonic decays of
charm and beauty hadrons
with an average
branching ratio to electrons of 10\% and 11\%, respectively.

The left panel of Fig.~\ref{DeBe} shows the momentum correlator versus relative azimuth for
$D-e$--pairs~(triangles), with the electron stemming from the semi-leptonic decay of
one of the $D$--mesons and for $e^{+}-e^{-}$--pairs~(squares) where both $D$--mesons
decayed into an electron.
The right panel shows the correlator for $e^{+}e^{-}$--pairs from semi-leptonic decays of $B\overline{B}$--mesons pairs.  
The correlations at small values of  $\Delta\phi$ do not survive the semi-leptonic decay while
at backward angles around $\Delta\phi\approx 180^{\rm o}$, we still observe a strong correlation.
We checked that this is due to the decay kinematics with gluon splitting processes dominating at forward angles resulting in a softer distribution of the 
heavy-quark hadron. At backward angles flavor creation processes dominate leading to a significantly harder spectrum.

Integrating the correlator over all azimuth and full rapidity,  we extract
$\corr = 0.007\pm0.001$~GeV$^2$/$c^2$  ($0.10\pm0.01$~GeV$^2$/$c^2$)  
for $e^{+}$-$e^{-}$--pairs from charm (bottom) decays
corresponding to the normalized dynamical 
fluctuation $\Sigma_{pt} \approx 17\% (18\%)$ with $\overline{p_t}$ = 0.50~GeV/$c$ (1.73~GeV/$c$).
This clearly indicates that the initial correlations among a 
heavy quark and its corresponding anti--quark even survive 
semi-leptonic decays into electrons (positrons) to a large extent.
With the predicted charm production in full rapidity of 0.16 per minimum bias  \pp collision at $\sqrt{s}$ = 14 TeV~\cite{ALICE_PPRII}, 
we  estimate the number of electron-positron pairs in the ALICE central barrel ($|y| < 0.9, p_t > 0.2$~GeV/$c$)  
from heavy-quark decays within one nominal year of ALICE running
to be more than 100k. Thus, an experimental observation of heavy-quark momentum correlation at the LHC should be possible. 

%
%
\begin{figure}[htb]
\centering
\includegraphics[width=0.40\textwidth]{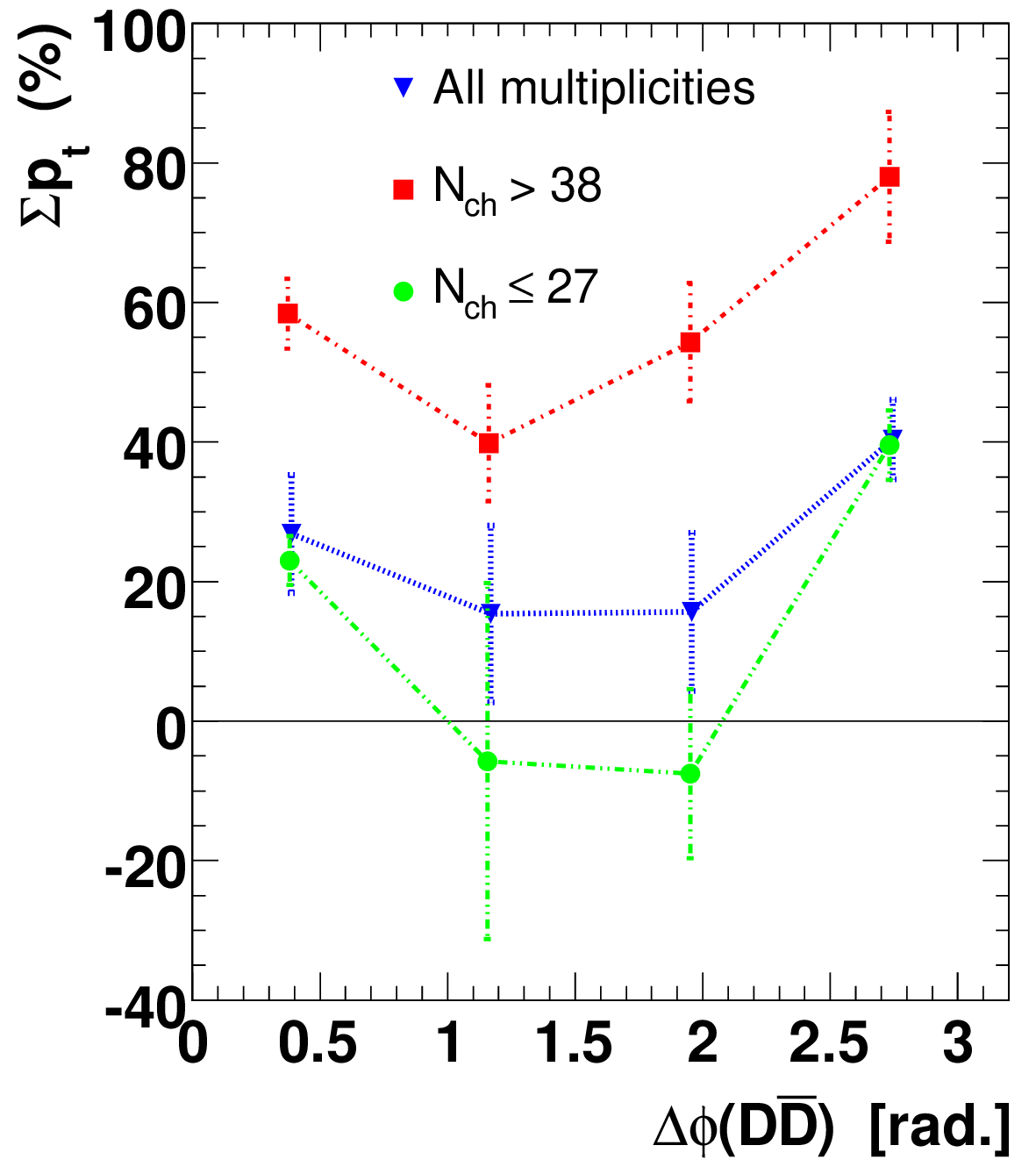}
\caption{(Color online) 
The normalized dynamical 
fluctuation  $\Sigma_{pt}$
 of $D\overline{D}$-pairs as function of 
$\Delta\phi$ for different multiplicity classes at full rapidity
from $pp$ collisions at $\sqrt{s}$ = 14 TeV as calculated using PYTHIA (v. 6.406). For each multiplicity class, 10k $D\overline{D}$-pairs were used. The lines connect the points.}
\label{mult}
\end{figure}

The heavy-quark transverse momentum depends on the relative contributions from different QCD-process. Furthermore, experimentally it has been observed that the average transverse momentum
is monotonically rising with the charged-particle multiplicity in \pp collisions~\cite{UA196,CDF2001}.
We studied the normalized dynamical 
fluctuation  $\Sigma_{pt}$
of $D\overline{D}$--pairs in several multiplicity regions
in \pp collisions at the top LHC energy as shown in Fig.~\ref{mult}. 
Higher multiplicity collisions result in stronger correlations due to the increase of the mean 
transverse momentum from
$\overline{p_t} = 1.53$ GeV/$c$ at multiplicities $\langle N_{ch} \rangle \approx 25$ to $\overline{p_t} = 3.46$ GeV/$c$ 
at $\langle N_{ch} \rangle \approx 42$.
In addition, the contribution from flavor creation to the production of $D$--mesons increases up to 18\%
in high multiplicity events compared with 12\% in low multiplicity events leading to 
enhanced correlations at large relative azimuth $\Delta\phi$. Thus, the highest multiplicities
in \pp collisions at LHC energies might be a good case to experimentally establish
the existence of these correlations for heavy quarks. 

The results on the heavy-quark correlator discussed above are our prediction for \pp collisions at the top LHC energy and serve as a baseline
for the case that no thermalization sets in as is expected for such a small collision system. 
Further, our calculations show that higher multiplicity $pp$ collisions result in stronger correlations. 

\begin{figure}[htb]
\centering
\includegraphics[width=0.4\textwidth]{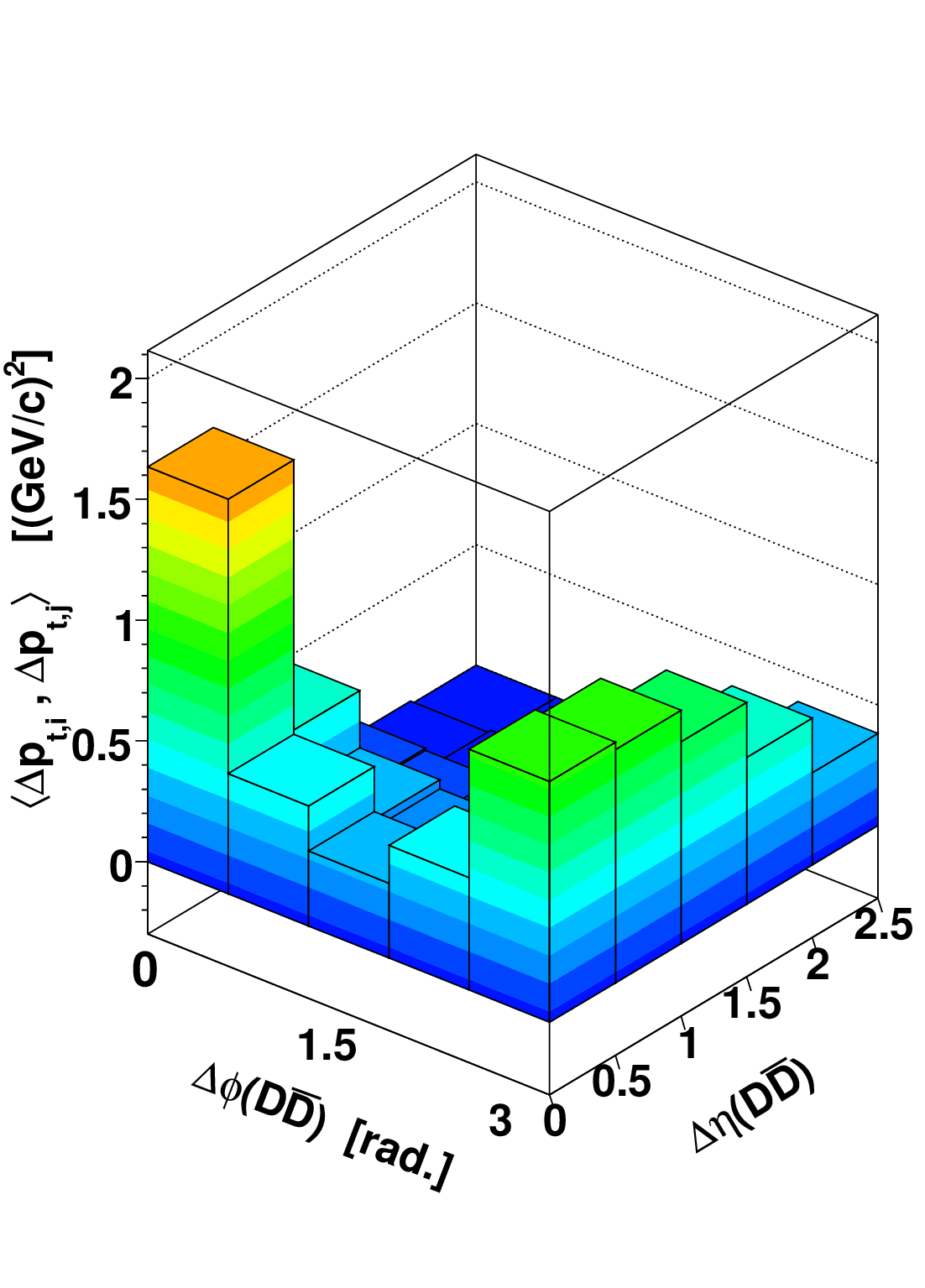}
\caption{(Color online) 
The momentum correlator of 2 million  $D\overline{D}$-pairs as a function of $\Delta\phi$  and $\Delta\eta$
from $pp$ collisions at $\sqrt{s}$ = 14 TeV as calculated using PYTHIA (v. 6.406).}
\label{eta_phi}
\end{figure}

Finally, we consider the relative pseudo-rapidity $\Delta\eta$ of $D\overline{D}$ pairs 
and show the correlator as a function of $\Delta\phi$  and $\Delta\eta$ (see Fig.~\ref{eta_phi}).
Gluon splitting processes lead to
correlations at small values of $\Delta\phi$ and $\Delta\eta$. On the other hand, flavor creation
results in correlations at large values of $\Delta\phi$ extending over a large range in $\Delta\eta$.
In  \pbpb\ collisions, the development of strong transverse flow would lead to a broadening of the away-side momentum correlation and an enhancement at
small relative azimuth of $D\overline{D}$ pairs~\cite{ismd08,molnar07}. 
To experimentally disentangle these different contributions, an analysis in ranges of pseudo-rapidity, e.g.  $|\Delta\eta| < 0.5$ to study
gluon splitting processes and effects of collective flow versus $|\Delta\eta| \ge 0.5$ where flavor creation dominates,    
might help.   

Further, a comparison of experimental heavy-quark correlations from \pbpb\ collisions to results from microscopic transport calculations 
would provide an independent way to extract effective heavy-quark scattering rates in the QCD-medium~\cite{molnar07}. 
Thus, it might be possible to extract general transport properties, which in turn provide important insight into the microscopic in-medium properties
of partons in the QGP and thus the nature of the plasma itself.  However, this information would be lost in case heavy quarks fully equilibrate with 
the light partons in the medium.

%

\section{Conclusions and outlook}
In summary, we have presented a sensitive method to see azimuthal correlations of heavy-quarks in $pp$ collisions at LHC energies.
We applied the momentum correlator to pairs of heavy-quark hadrons and their semi-leptonic decay products 
as a precise and normalized measure. At LHC energies, the production of charm quarks is expected to be dominated by gluon splitting
processes resulting in forward correlations with an increasing contribution of backward-peaked pair creation at larger momentum. A stronger correlation is expected in high-multiplicity $pp$ collisions.  
A modification or disappearance of these momentum correlations in \pbpb\ collisions as compared to \pp collisions
will be explored as a sensitive probe of thermalization.  Also, cold-nuclear matter effects on heavy--quark correlations should be studied in $p(d)+A$ collisions, serving as another important baseline.

 On the other hand, open heavy-quark correlations also become important in the study of dilepton invariant-mass spectra
in high-energy nuclear collisions, since the contribution of correlated open heavy-quark decays competes with dilepton emission rates
from the QGP in the intermediate invariant-mass range~\cite{ceres02, PHENIX2010}. Thus, solid experimental constraints on the extent of the loss of
open heavy-quark correlations are essential for an interpretation of the dilepton invariant-mass spectra, in particular concerning mechansims 
of chiral symmetry restoration in the hot partonic medium created in high-energy nuclear collsions at LHC energies.


\section{Acknowledgment}
After re-submission of this manuscript, the authors learned that there exists a publication on azimuthal correlations of leptons stemming from
semi-leptonic decays of heavy quarks probing the QGP~\cite{hirano}.
We would like to thank J.~Castillo, Dr.~Y.~Pachmayer, Dr.~D.~Miskowiec, and Dr.~N.~Xu  for exciting discussions.
This work was supported by the Helmholtz Association under contract number VH-NG-147
 and the Helmholtz Alliance HA216/EMMI.


\begin{thebibliography}{99}


\bibitem{strange} 
J. Adams {\it et al.} (STAR collaboration), Phys. Rev. Lett. {\bf 95}, 122301 (2005); 
B.I. Abelev {\it et al.} (STAR collaboration), Phys. Rev. Lett. {\bf 99}, 112301 (2007).  
\bibitem{thermalization}
B. Svetitsky, Phys. Rev. D {\bf 87}, 2484 (1988).
\bibitem{quark-masses}
J. Gasser and H. Leutwyler,  Phys. Rept. {\bf 87}, 77 (1982). 
\bibitem{pbm_2007} 
A.~Andronic, P.~Braun-Munzinger, K.~Redlich, and J.~Stachel, Nucl. Phys. {\bf A 789}, 334 (2007);  
P.~Braun-Munzinger,~J.~Phys.~G~{\bf 34},~S471~(2007).
\bibitem{aleph_95}  B. Buskulic  {\it et al.} (ALEPH Collaboration) Phys.~Lett.~B {\bf 357}, 699 (1995).
\bibitem{delphi_95} P. Abreu  {\it et al.} (DELPHI Collaboration) Z. Phys. C {\bf 66}, 323 (1995).
\bibitem{opal_95} G. Alexander  {\it et al.} (OPAL Collaboration) Phys.~Lett.~B {\bf 364}, 93 (1991).
\bibitem{review}C. Louren\c{c}o and H.K. W\"ohri, Phys. Rept. {\bf 433}, 127 (2006). 
\bibitem{e791}
E.M. Aitala  {\it et al.} (E791 Collaboration), Eur. Phys. J. direct C {\bf 1}, 1 (1999).
\bibitem{Zhu} 
X.~Zhu, M.~Bleicher, S.L.~Huang, K.~Schweda, H.~St\"ocker, N.~Xu, and P.~Zhuang, 
Phys. Lett. B {\bf 647}, 366 (2007).
\bibitem{hq-thermalization}
see for example: \\
H. van Hees, V. Greco, R. Rapp, Phys. Rev. C {\bf 73}, 034913 (2006); \\
G.D. Moore, D. Teaney, Phys. Rev. C {\bf 71}, 06904 (2005);\\
P. Petreczky, D. Teaney, Phys. Rev. D {\bf 73} 014508 (2006).
\bibitem{pythia}T. Sj\"ostrand {\it et al.}, Comput. Phys. Commun. {\bf 135}, 238 (2001). 
\bibitem{ALICE_PPRII} B.Allesandro {\it et al.} (ALICE Collaboration), 
 J. Phys. G {\bf 32}, 1295 (2006); we used the values for  the PYTHIA parameters as listed in Table 6.54.
\bibitem{pythia2}E. Norrbin and T. Sj\"ostrand, Eur. Phys. J. C {\bf 17}, 137 (2000). 
\bibitem{Zhu2} X.~Zhu, N.~Xu, and P.~Zhuang, Phys. Rev. Lett. {\bf 100}, 152301 (2008).
\bibitem{ptcumul}A.~Bialas and M. Gazdzicki, Phys.~Lett.~B {\bf 252}, 483 (1990).
\bibitem{na49PRL}T. Anticic {\it et al.} (NA49 Collaboration), Phys. Rev. C {\bf 70} 034902 (2004).
\bibitem{koch99} S.A. Voloshin, V. Koch, and H.G. Ritter, Phys. Rev. C {\bf60}, 024901 (1999).
\bibitem{ceres-pt} D. Adamova {\it et al.} (CERES collaboration), Nucl. Phys. A {\bf 811}, 179 (2008).
\bibitem{spt}    D. Adamova {\it et al.} (CERES collaboration), Nucl. Phys. A {\bf 727}, 97 (2003).
\bibitem{spt1}H. Sako {\it et al.} (CERES collaboration), J. Phys. G {\bf 30}, S1371 (2004).
\bibitem{spt2}M. Rybczynski {\it et al.} (NA49 collaboration), J. Phys. Conf. Ser.  {\bf 5}, 74 (2005).
\bibitem{spt3} J. Adams {\it et al.} (STAR collaboration), Phys. Rev. C {\bf72} 044902 (2005).
\bibitem{sqm08}K. Schweda and G. Tsiledakis, J. Phys. G {\bf 36},  064058 (2009).
\bibitem{UA196} G.~Bocquet  {\it et al.} (UA1 collaboration),  Phys.~Lett.~B {\bf 366}, 434 (1996).
\bibitem{CDF2001} F.~Abe  {\it et al.} (CDF collaboration), Phys. Rev. Lett. {\bf 61}, 1819 (1988).
\bibitem{ismd08}G. Tsiledakis and K. Schweda, {\it proc. of the ISMD08 conf.}, DESY-PROC-2009-001, 214 (2009).
\bibitem{molnar07}D. Molnar, Eur. Phys. J. C {\bf49}, 18 (2007).
\bibitem{ceres02} D. Adamova {\it et al.} (CERES/NA45 collaboration), Phys. Rev. Lett. {\bf 91},  042301 (2003). 
\bibitem{PHENIX2010} A. Adare  {\it et al.} (PHENIX collaboration),  Phys. Rev. C {\bf 81},  034911 (2010). 
\bibitem{hirano}Y. Akamatsu, T. Hatsuda, and T. Hirano, Phys. Rev. C {\bf 80},   031901(R)  (2009).  
\end{thebibliography}
\end{document}